\documentclass[prb,preprint]{revtex4-1}

\usepackage{amsmath}  
\usepackage{amsfonts} 
\usepackage{graphicx}
\usepackage{verbatim}

\begin{document}

\title{Simulating  gravitational motion, gas dynamics, and structure  in the cosmos}

\author{J. W. Powell}
\email{dna@reed.edu}
\affiliation{Department of Physics, Reed College, Portland, Oregon 97201}
\author{L. Caudill}
\affiliation{Department of Physics, Reed College, Portland, Oregon 97201}
\author{O. Young}
\affiliation{Department of Physics, Reed College, Portland, Oregon 97201}
\date{\today}

\begin{abstract}
We provide introductory explanations and illustrations of the $N$-body hydrodynamics code Charm N-body GrAvity solver (ChaNGa). ChaNGa simulates the gravitational motion and gas dynamics of matter in space, with the goal of modeling galactic and/or cosmological structure and evolution.  We  discuss  the algorithm for leapfrog integration and smoothed particle hydrodynamics and computer science concepts used by the program, including the  binary data structure for the particle positions. Our presentation borrows  from the  doctoral dissertation of J.\ G.\ Stadel. Problems are provided in order to use ChaNGa to learn
or solidify some cosmological concepts. 
\end{abstract}

\maketitle

\section{Introduction}

N-body hydrodynamic simulations of the cosmos and isolated galaxies\cite{springel-2005, hopkins-2013}   have been crucial in establishing  Lambda cold dark matter (Lambda refers to dark energy) as the standard model of cosmology. \cite{Frenk-and -White-2012}  Due to the fundamental nature of gravity, computationally challenging  densities occur in astrophysical situations. Simulations developed by several groups have  yielded trustworthy  predictions of the   incredibly diverse structure of the cosmos and galaxies, and   several  simulations of very differing kinds have given the same results, thanks to the AGORA project.\cite{agora-II}

ChaNGa (Charm N-body GrAvity solver) is an example of a highly parallel N-body hydrodynamic code.\cite{menon}  The physics of star formation, supernovae feedback, and cooling are included in ChaNGa.

Our goal is to provide    an accessible, but thorough explanation of how to use this  N-body hydrodynamic code by adding to ChaNGa's GitHub wiki~\cite{wiki1} and many YouTube presentations. We also hope to provide a deeper appreciation for how computer science  data structures are used to construct  ChaNGa.  For example,  ChaNGa is written in the parallel programming language CHARM++, which won a Gordon Bell prize.\cite{prize} Although we have not attempted to provide an exhaustive description of the code, we focus on the details that are most relevant to the physics.

Sections~\ref{sec:code gravity}--\ref{sec:barnes-hut changa} discuss gravitational motion and the ChaNGa's approach to treating it. Section~\ref{sec:boundary conditions} describes several considerations for a  simulation of an infinite (or at least very large) universe. Section~\ref{sec:force softening} describes a technique for softning gravitational forces at small distances to avoid problems that would otherwise arise in the calculation of the forces. Section~\ref{sec:sph} describes the method by which ChaNGa accounts for the presence of gas in the universe.  Section~\ref{sec:results} gives some of the results of ChaNGa and Sec.~\ref{sec:exercises} provides some suggested problems.

\section{Newtonian Gravity and Numerical Integration}\label{sec:code gravity}

Gravity is responsible for Kepler's three laws which are commonly derived in junior-level mechanics without any mention of computers. The difference between the $N = 2$ and $N > 2$  systems is  huge despite the fact that Newton's laws and the universal law of gravitation are the same for both.
	
The gravitational force between two massive particles $m_1$ and $m_2$ in three dimensions is given by
	\begin{equation}\label{eq:gravity vec}
	    \vec{F}_1=-\vec{F}_2=\frac{Gm_1m_2}{|\vec{r}_2-\vec{r}_1|^3}(\vec{r}_2-\vec{r}_1),
	\end{equation}
where $\vec{r}_1$ and $\vec{r}_2$ denote the position vectors of the two particles.  Codes capable of treating the N-body problem relativistically do not currently exist.\cite{finnish} Because the forces are additive,   the gravitational force acting on the $i$th particle in a system of three or more  particles is
	\begin{equation}\label{eq:gravity vec sum}
	    \vec{F}_i=\sum_{j\neq i}^{N}\frac{Gm_im_j}{|\vec{r}_j-\vec{r}_i|^3}(\vec{r}_j-\vec{r}_i).
	\end{equation}
Equation~(\ref{eq:gravity vec sum}) gives  the Newtonian approximation to gravity. We need only substitute Eq.~\eqref{eq:gravity vec sum}  into Newton's second law  and solve the resulting  second-order differential equation for all the particles  to model a galaxy. The resulting equation has no analytical solutions\cite{beckett} for $N>2$. Rather than trying to derive the exact trajectories for $N\gg2$ particles, we need to use a numerical integration technique to approximate the trajectories.

A  common numerical method  for force integration for N-body codes,\cite{quinn} which  is similar to the method adopted by ChaNGa,\cite{beckett} is known as leapfrog integration. Any numerical integration algorithm requires dividing the time into discrete steps. The algorithm  for leapfrog integration (derived by a Taylor expansion) is\cite{beckett}
\begin{align}
\label{eq:leapfrog0velocity}
\vec{v}_{n+\frac{1}{2}}&=\vec{v}_{n-\frac{1}{2}}+\vec{a}_n\Delta t\\
\label{eq:leapfrog0position}
\vec{r}_{n+1}&=\vec{r}_n+\vec{v}_{n+\frac{1}{2}}\Delta t.
\end{align}
The form of Eq.~\eqref{eq:leapfrog0velocity} suggests the etymology of ``leapfrog''\cite{quinn}:  given initial values for the position and velocity,  we first use Eq.~(\ref{eq:leapfrog0velocity}) to update the velocity, and then  use the new velocity in 
	Eq.~(\ref{eq:leapfrog0position}) to update the position. We    repeat the process many times, alternating between velocity and position in a manner similar to the game ``leapfrog.'' 
	
Although Eqs.~(\ref{eq:leapfrog0velocity}) and (\ref{eq:leapfrog0position}) give  the algorithm in a  compact form and provide insight into a  familiar technique, they are  not the equations that ChaNGa actually uses in its implementation. Instead, ChaNGa uses what is known as the ``kick-drift-kick'' form of the algorithm, which can be written as\cite{beckett,mitch}
\begin{align}\label{eq:kick drift kick 1}
\vec{v}_{n+\frac{1}{2}} & =\vec{v}_n+\vec{a}_n \frac{\Delta t}{2} \\
\label{eq:kick drift kick 2}
\vec{r}_{n+1} & =\vec{r}_n+\vec{v}_{n+\frac{1}{2}}\Delta t \\
\label{eq:kick drift kick 3}
\vec{v}_{n+1}& =\vec{v}_{n+\frac{1}{2}}+\vec{a}_{n+1}\frac{\Delta t}{2}.
\end{align}
From the form of Eqs.~\eqref{eq:kick drift kick 1}--\eqref{eq:kick drift kick 3}, we  see where the name ``kick-drift-kick'' arises. At each step, we first update the intermediate velocity term $\vec{v}_{n+\frac{1}{2}}$ using the current gravitational acceleration (equal to the force per unit mass acting as the kick), then update the position $\vec{r}_{n+1}$ according to this new velocity (Eq.~\eqref{eq:kick drift kick 2}  treats the particle as though it is moving in the absence of external forces, i.e., drifting), and then update the velocity $\vec{v}_{n+1}$ using the gravitational acceleration at the new spatial position. 
	
There are several reasons why the kick-drift-kick form of the leapfrog algorithm is favored by ChaNGa. Paramount among these is a concept known as  \emph{multi-stepping}.\cite{menon} The number of computations in Eqs.(\ref{eq:kick drift kick 1})--(\ref{eq:kick drift kick 3})  made during a fixed time interval  largely  determines the  accuracy of this integration algorithm. Because the method  is  an approximation in which particles move with a fixed velocity between time steps, the smaller the time step, the more accurate the trajectories. There is a  tradeoff in terms of run time, and so rather than pursuing increasingly small time steps, the goal is usually to find the number of time steps  compared to the duration of the time interval for which the  accuracy is good enough. Because gravity is  spatially dependent,  what constitutes sufficient temporal resolution is closely linked with the particle density. Regions with higher particle density will naturally require smaller time steps  to achieve the same level of physical accuracy that can be attained in a lower density region with a larger time step.\cite{beckett,mitch,springel-2005}

One approach to this problem would be to increase the number of time steps globally to be suitable to the desired level of accuracy in high density regions, but doing so would introduce a large number of  extraneous calculations for low density regions and thus would  be computationally wasteful.\cite{springel-2005} The solution  is to assign time steps on a per-particle basis; that is, each particle gets its own time step  in accordance with the density of its surrounding particles (multi-stepping). The kick-drift-kick method ensures that such an individualized time-step assignment is possible, as long as the time step sizes of all particles are related by powers of two.\cite{beckett,stadel-time-stepping}

\section{A Hierarchical Data Structure for the Particle Positions: The Barnes-Hut Algorithm}\label{sec:barnes-hut}
	
Due to the complex nature of the N-body problem, another challenge is    run time efficiency. To understand the challenge,  note that the net acceleration  in the leapfrog algorithm in Eqs.~(\ref{eq:kick drift kick 1})-- (\ref{eq:kick drift kick 3}) is  the result of the   large sum of gravitational interactions  in Eq.~(\ref{eq:gravity vec sum}) for    $ N \gg 1$.  From the form of  Eq.~(\ref{eq:gravity vec sum}), we see that for a system of $N$ particles, each particle experiences $N-1$ gravitational interactions. If we were to write a  program with just the information from Sec.~\ref{sec:code gravity},  our program would need to perform $N-1$ force calculations $N$ times or $\sim N^2$ calculations at each time step.\cite{finnish}    This number might be acceptable for small-scale computations, say for $N=100$ to $N=1000$,  but the current state of computational astrophysics is such that simulations  involve $N$ well in the millions and often evolve over thousands of time steps.\cite{beckett,mitch} Programs need to accommodate  this level of scale, which requires methods to speed up the gravitational force calculation process. 

The solution employed by ChaNGa is to use a data structure designed to speed up the access to the huge amount of initial positions data: a tree structure. The most well known implementation of such a structure is the Barnes-Hut  algorithm.\cite{barnes-hut}  
Barnes-Hut offers an approximation of the gravitational force acting on a particle in logarithmic time ($O[\log N)$], making the total force calculation per time step $O(N\log N)$, a  gargantuan improvement from the polynomial time direct summation method.

In the following, we introduce the standard Barnes-Hut algorithm   to familiarize the reader with the general approach to the Gordon-Bell-prize-winning solution for speeding up the N-body problem. We will then   discuss some of the modifications employed by ChaNGa and the motivations behind them. Readers who wish  a more in-depth or specific treatment of Barnes-Hut might be interested in  the seminal 1986 paper,\cite{barnes-hut} or  more accessible sources such as Refs.~\onlinecite{barnes-hut-princeton} and \onlinecite{beckett}.

The fundamental principle behind the Barnes-Hut algorithm is to reduce the number of  force calculations performed for each particle by applying a center-of-mass approximation for the force exerted by collections or clusters of distant particles.\cite{demmel} More explicitly, for distant clusters -- particles that are  far from the particle whose acceleration we want to calculate but are  close to each other\cite{demmel} -- the algorithm determines the center of mass and total mass of the cluster, and then performs a  single gravitational force calculation  instead of many force calculations which contribute little to improving the accuracy of the total force. 
	
As noted in Ref.~\onlinecite{beckett}, the nature of gravitation is such that for a sufficiently distant cluster of identical particles, the force contribution associated with each particle will be roughly the same (both in  magnitude and direction), so the approximation is justified. Of course, distance constitutes the main underlying assumption for the approximation, meaning it does not hold for nearby particles, for which direct summation is still performed. 	
Because this approach relies  heavily on distance, we need to establish a way to determine what constitutes ``far enough'' for a center-of-mass approximation and ``close enough'' for particles to be considered part of the same cluster. We might be tempted to simply compare the distances of all the particles in the simulation, but such a process would be $O(N^2)$,\cite{beckett} 
and be no better than direct summation.

The Barnes-Hut algorithm is a  clever solution to this problem, which starts with a process known as  \emph{domain decomposition}, or the repeated decomposition of the initial volume into smaller subvolumes. It is useful to define the widely used term  \emph{node} in the context of hierarchical data structures:  ``each node represents a portion of the 3-D space containing the particles in that volume.''\cite{menon}   The  Barnes-Hut algorithm uses an Octree decomposition data structure,\cite{beckett,barnes-hut} which involves recursively dividing cubical volumes into eight subvolumes, hence the name. (This decomposition is not the  scheme ChaNGa uses, but understanding the basic Octree decomposition method is helpful as an introduction to the method in ChaNGa.) A summary of Octree decomposition is as follows:
\begin{enumerate}
        \item Begin with the volume containing all particles, this volume is the \emph{root node} of the Octree.
        
        \item Make the first decomopostion by dividing the root node evenly into eight subvolumes.
        
        \item If there are no particles in one of the subvolumes/nodes, discard it.  
        
        \item If two or more particles are present in a subvolume, divide it evenly into eight daughter subvolumes/nodes. Then inspect each daughter node individually.
        
        \item Repeat steps 2--4 until each node contains at most one particle.
\end{enumerate}
In total, the  complexity of constructing the Octree is $O(N\log N)$.\cite{barnes-hut}
  
\begin{figure}[h!]
\centering
\includegraphics[scale=0.3]{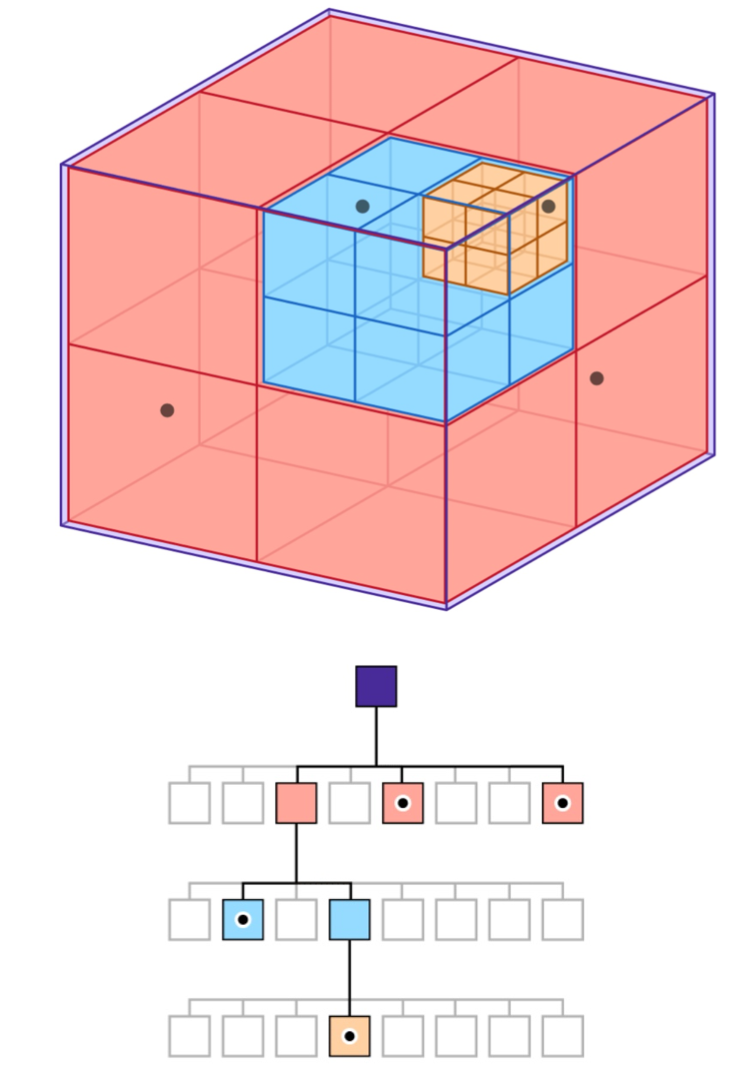}
\caption{(Color online) A visual representation of an Octree decomposition for four particles taken from Ref.~\onlinecite{apple}.  The figure on underneath gives an alternative visualization of the decomposition. The purple node at the top corresponds to the entire original volume. The next layer down  is  orange-pink, with    three kinds of boxes in that row: no color for  no particles, orange-pink with a black dot indicating a single particle, and  an orange-pink box with no black dots indicating that there are two or more particles in that node (two particles in this example). These nodes require further decomposition.}
\label{fig:octree}
\end{figure}

Figure~\ref{fig:octree} gives an example of this process for a system of  four particles. The nodes vary in size according to their corresponding level in the tree structure. Only boxes containing more than one particle are subdivided into  daughter nodes, and division is performed recursively until each node contains one or zero particles.  In practice, the number of particles which is the threshold for further decomposition is higher than one and is a parameter that can be varied to adjust the resolution.  See Ref.~\onlinecite{demmel} for an excellent  description of this process.
    
We still need to discuss how  this hierarchical data structure expedites the force calculation, an insight which may yet elude  all but the most computer-savvy readers. 
    
The next step involves center of mass calculations. The implementation of the algorithm as described in Ref.~\onlinecite{barnes-hut} is such that the data structure representing a node in the tree has attributes corresponding to the total mass and center of mass  for the particles which it contains. For nodes containing only one particle (``leaf nodes'' in the language of the tree metaphor), these values are trivial, so it is most efficient to propagate this information backward through the data structure, i.e., leaves-to-root, a process which is also $O(N\log N)$.\cite{barnes-hut} 
    
\begin{figure}[t]
\centering
\includegraphics[scale=0.7]{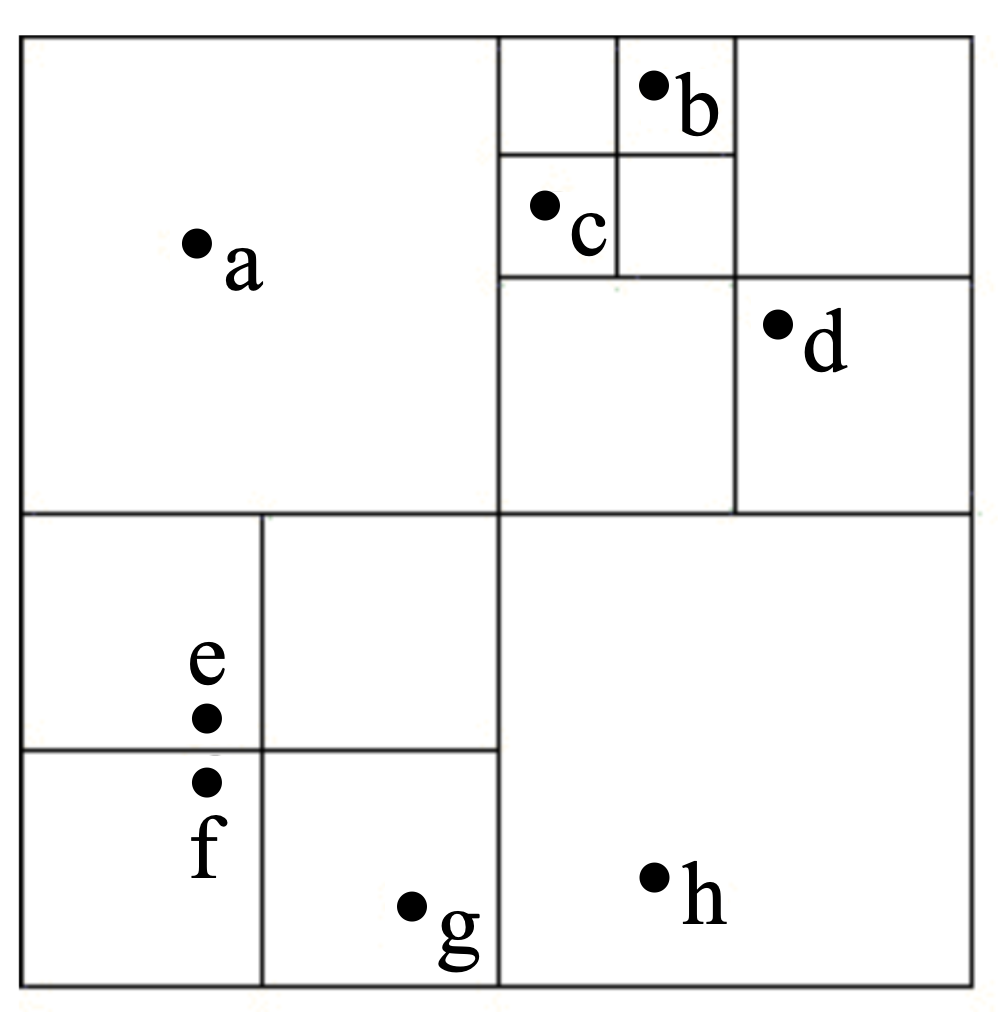}
\caption{A two-dimensional analogue of the Barnes-Hut Octree, known as a Quadtree.  Such figures are better understood  initially by removing all the subvolumes and starting from scratch.  Only domain decomposed subvolumes that have more than one particle are shown in this figure. Figure 3 shows an alternate interpretation of the same information.  This image was taken from  Ref.~\onlinecite{princeton}.}
        \label{fig:quadtree-1}
\end{figure}

\begin{figure}[t]
\centering
\includegraphics[scale=0.7]{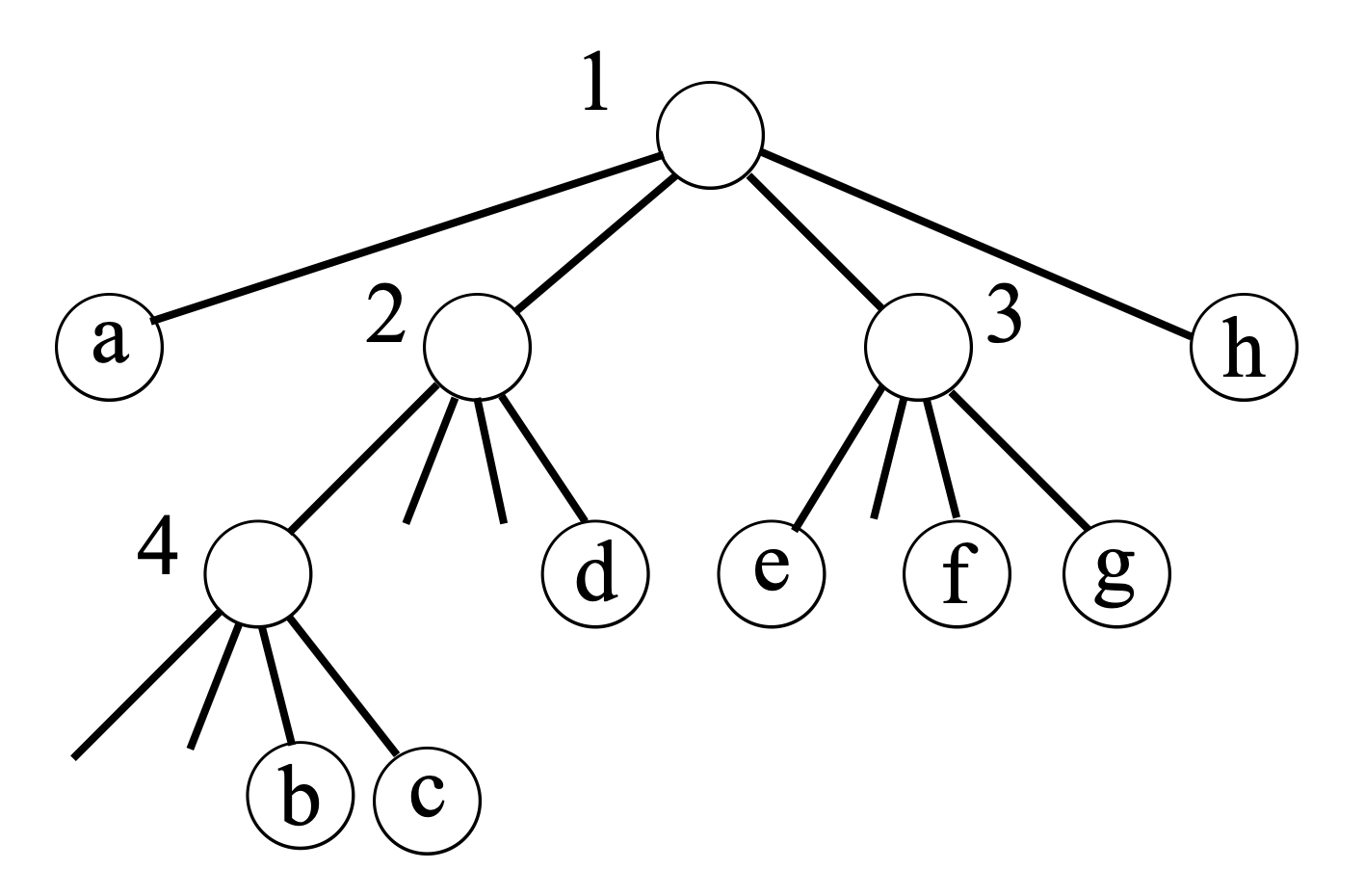}
\caption{An alternate image of Fig. 2., two-dimensional analogue of the Barnes-Hut Octree, known as a Quadtree.  Such figures are better understood  initially by removing all the subvolumes and starting from scratch.  The numbers next to the circles on the this figure  refer to a particular node in the text. This image was taken from  Ref.~\onlinecite{princeton}.}
        \label{fig:quadtree-2}
\end{figure}

If we examine the analogous two-dimensional Quadtree data structure depicted in Figs.~\ref{fig:quadtree-1} and ~\ref{fig:quadtree-2}, we  see that the center of mass and total mass calculation would begin with node 4, whose  attributes would be calculated using the positions and masses of particles a and b. We can then use these attributes of node 4, as well as the mass and position of particle d, to compute the total mass and center of mass of node 2 and so on. Nodes containing more than one particle can then be treated as \emph{pseudoparticles} with their own associated mass (the total mass of the particles)  and position (the center of mass position).
    
Once all the  pseudoparticle nodes in the system have been assigned their mass and center of mass attributes, all the machinery necessary to begin force calculations is in place. The algorithm for calculating the force on a given particle $p$ involves a traversal of the Octree, where $\ell$ is the side length of the cubical region represented by the current node, $D$ is the distance between $p$ and the node's center of mass, and $\theta$ is an accuracy parameter set at the start of the simulation, usually $\sim 1$. The algorithm for each particle is\cite{barnes-hut}
\begin{enumerate}
        \item Start at the root node. 
        
\item If $\ell/D<\theta$, compute the gravitational force between the current node and $p$, and add it to the total force acting on $p$.
        
\item Otherwise, traverse one layer down the tree (away from the root node)  and perform step 2 for each daughter node.
        
\end{enumerate}
    
Once these steps have been completed, the gravitational influence of each particle on $p$ is   accounted for either by a direct force calculation or by a center of mass approximation. For large $N$, this process involves performing order $\log N$ force calculations for each particle, and thus the  overall run time of the algorithm is $O(N\log N)$.
    
\section{Barnes-Hut Modifications}\label{sec:barnes-hut changa}
	
The description we have provided  of the Barnes-Hut gravitational force computation closely follows Ref.~\onlinecite{barnes-hut}, but as  mentioned, there are several changes in the ChaNGa code. Most of these changes are described in Ref.~\onlinecite{stadel}.  One notable change is that ChaNGa does not actually use the center of mass of far away  clusters to approximate their gravitational influence. Instead, ChaNGa performs an operation known as a multipole expansion (usually discussed for electric fields)  for improved force accuracy.\cite{menon} See Ref.~\onlinecite{demmel} for an excellent description of multipole expansions for $N$-body simulations.  Also see  the discussion of multipole expansions in Ref.~\onlinecite{griffiths}. Due to the similarities between Newtonian gravity and the electric field, multipole expansions  can  be used to approximate the gravitational potential produced by clusters of massive particles.
	
A multipole expansion effectively amounts to an infinite sum of terms  with increasing angular dependence.\cite{multipole} The first (or zeroth order) term is a monopole, or a point source, which has no angular dependence. The first few higher order terms are the dipole, quadrupole, and hexadecapole terms  in the electromagnetism context,\cite{griffiths, multipole} or in  Ref.~\onlinecite{stadel} in the $N$-body simulation context. ChaNGa performs a multipole expansion to third order, including all terms up to and including hexadecapole,\cite{menon} which   is  faster and more accurate than a quadrupole order expansion.\cite{stadel}.
	
Another notable difference between the  Barnes-Hut algorithm and that employed by ChaNGa is the  precise structure of the spatial decomposition tree. As we discussed in Sec.~\ref{sec:barnes-hut}, the algorithm    recursively divides the spatial domain into even sections of eight, resulting in a data structure called an Octree (see Fig.~\ref{fig:octree}). Instead, the spatial decomposition performed by ChaNGa  employs a  binary tree rather than an Octree, meaning the domain is recursively divided into sections of two rather than eight. This change is consistent with the approach used by the $N$-body code PKDGRAV,\cite{stadel} from which ChaNGa inherits much of its gravitational force calculation.\cite{menon}

The justification for using a binary tree rather than an Octree 
is that a binary tree structure offers advantages for parallelization, particularly when distributing the computation over an arbitrary number of processors,\cite{stadel} a feature which has become of central importance in    high performance computing.
	
\begin{figure}[t]
\centering
\includegraphics[scale=0.7]{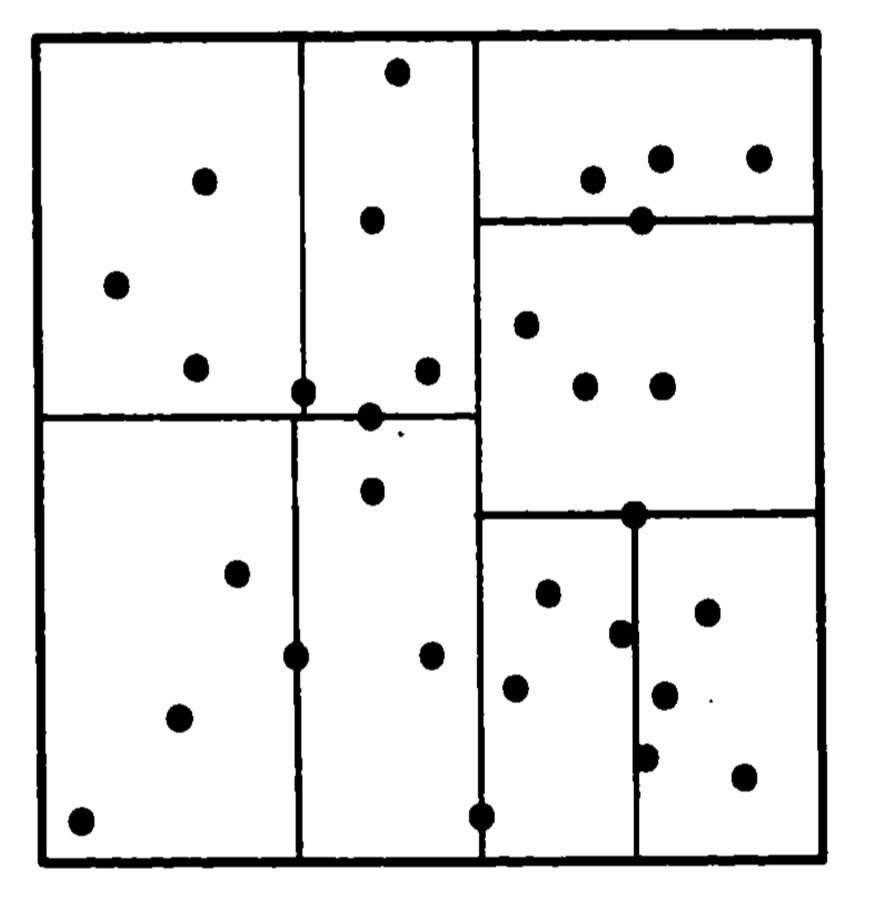}
\caption{An Image representing a standard k-D tree  domain decomposition.. The k-D tree is constructed with a maximum of four particles per leaf cell.  On the right the bounds have been squeezed before bisection. The dashed lines indicate bisectors. The differences between the two decompositions in Figs. 4 and 5 are much more pronounced for a more highly clustered distribution.  In that case the spatial locality of the spatial binary tree ( Fig. 5 ) would be much improved over the k-D tree.  k-D trees can have pathological problems with certain distributions,\cite{stadel} so the spatial binary tree ( Fig. 5 ) is the best choice.  This image and caption are  taken from Ref.~\onlinecite{stadel}}.
\label{fig:binary tree-1}
\end{figure}

\begin{figure}[t]
\centering
\includegraphics[scale=0.7]{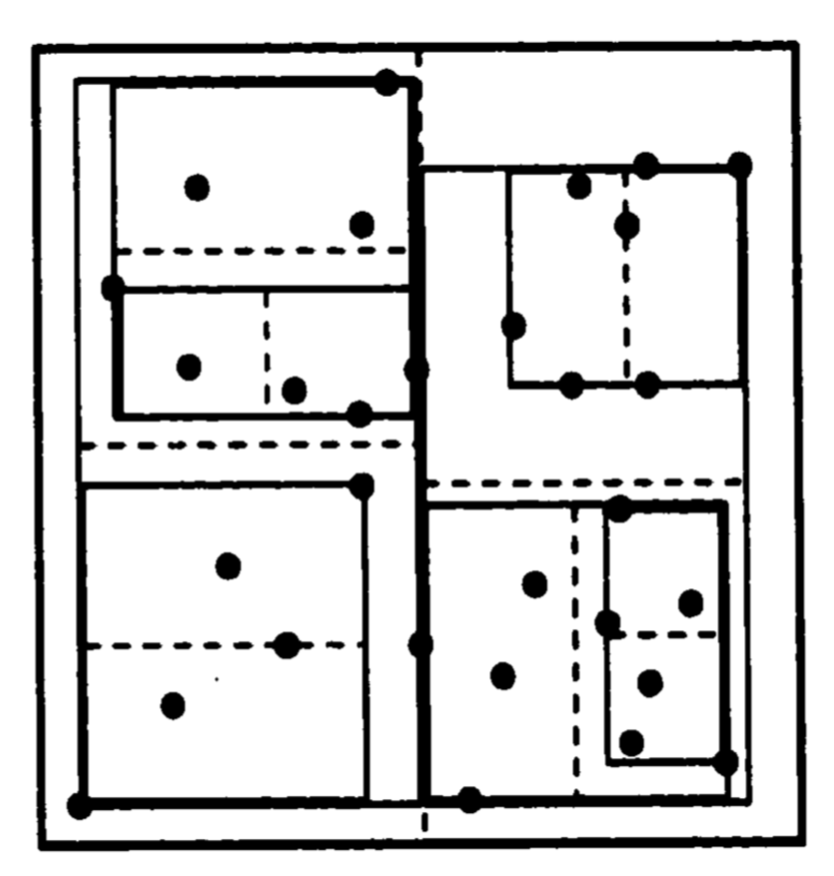}
\caption{An image representingthe spatial binary tree decomposition used by PKDGRAV.  This figure shows the bounds have been squeezed before bisection. The dashed lines indicate bisectors. The differences between the two decompositions is much more pronounced for a more highly clustered distribution.  In that case the spatial locality of the spatial binary tree would be much improved over the k-D tree.  k-D trees can have pathological problems with certain distributions,\cite{stadel} so the spatial binary tree is the best choice. The name spatial binary tree comes from ``spatially bisecting the bounding box.'' This image and caption are  taken from Ref.~\onlinecite{stadel}}.
\label{fig:binary tree-2}
\end{figure}

Although the precise details of ChaNGa's spatial binary tree implementation are not easy to learn from      the literature, ChaNGa explicitly inherits much of its gravitational force calculation from PKDGRAV,\cite{menon} the details of which can be found in Ref.~\onlinecite{stadel}. Because of this intellectual inheritance, a brief discussion on PKDGRAV's domain decomposition method is warranted to provide a better understanding of what is implemented in ChaNGa. Aside from the overall difference in organizational structure, a crucial difference between the Barnes-Hut Octree and the binary trees used by contemporary simulation codes is that in binary tree data structures space is often  not divided up evenly. Instead, daughter nodes are sized dynamically in accordance with some bisection scheme, often related to the number of particles per daughter node. An example of such a binary tree algorithm (in two dimensions) is the \emph{k-D tree}, depicted  Fig.~\ref{fig:binary tree-1}. This algorithm involves recursively bisecting nodes through their longest axis such that both daughter nodes contain approximately the same number of particles. Note that in Fig.~\ref{fig:binary tree-1}  the first bisection is done such that each daughter node contains 15 particles, then 8, and finally 4.  
	
The k-D tree offers a good first look into spatial binary trees  with dynamically sized nodes, but it also presents problems in terms of force error and run time efficiency.\cite{stadel} The decomposition method used by PKDGRAV  is depicted in two dimensions in Fig.~\ref{fig:binary tree-2}. By comparing Fig. 5 to Fig. 4 representing the k-D tree, it is clear that there is more going on here than explained in the caption.  You can gain a better understanding of spatial binary trees by reading the appropriate section of Ref.~\onlinecite{stadel}.  
	
To better understand what is shown in Fig.~\ref{fig:binary tree-2}, we  note that the spatial binary tree decomposition  really has only one fundamental difference from the k-D tree: the ``squeezing'' of daughter nodes to minimize the volume they represent. At every step, PKDGRAV's binary tree compresses the volume of the node being examined to the smallest rectangle (rectangular prism in three dimensions) that contains all the particles in the node. Because the bounding box is now considered to be the node's volume, it is bisected through its longest axis into two daughter nodes containing roughly equal numbers of particles,  which are then squeezed themselves. This process is repeated until each node is under the maximum particle count for a leaf node, which for ChaNGa is usually around 8--12.\cite{menon}
	
We have so far discussed two of the fundamental modifications to the Barnes-Hut algorithm employed by ChaNGa, namely, the use of a hexadecapole order multipole expansion for the gravitational force approximation at large distances, and the use of a spatial binary tree structure rather than an Octree. There is another aspect of the gravitational force calculation that   should be mentioned,  namely,  ChaNGa's approach to parallelization, because large-scale parallelization is a primary raison d'\^{e}tre for the code.\cite{massively-parallel} A detailed description of the parallelization process can be found in Sec.~4 of Ref.~\onlinecite{menon}.  
	
We have described how  ChaNGa uses a spatial binary tree to handle the gravitational force calculation. Although it is correct that the binary tree is the fundamental data structure, ChaNGa's force calculation  involves not one but a number of spatial binary trees, each representing a subregion of the simulation volume containing a subset of the overall particle count, divided  among the processors allotted to the computation.\cite{menon} To facilitate this approach, ChaNGa performs an initial domain decomposition step not described by the spatial binary tree or Octree, in which the simulation volume is divided into a number of subregions containing an equal number of particles according to a  space-filling curve  algorithm.\cite{menon}  The  space-filling curve passes through each discrete spatial cell containing a particle. The resulting spatial decomposition looks different from a binary tree or Octree.  This space-filling curve algorithm initially occurs  to balance the computational load across the processors being used. The code iterates through a number of potential decompositions until all bins (subregions) are  sufficiently optimal.\cite{menon} All the particles in each bin are then assigned to a data object called a  \emph{tree piece},   so that each tree piece represents a subset of the overall volume. Each tree piece then performs a spatial binary decomposition of its assigned subvolume (using a bounding box method similar or identical to the one we have  described), and the gravitational force calculation begins.\cite{menon}

The term ``tree piece'' is closely associated with another concept in ChaNGa; a  \emph{chare}. Chares are important in forming  checkpoints for long simulations, and chares are essentially  tree pieces.\cite{menon} Note that, because particles are assigned to tree pieces in accordance with the initial space-filling curve decomposition, no single processor has direct access to all the particles in the simulation, but all particles interact with each other gravitationally. Because of this dichotomy, ChaNGa also facilitates communication across processors to allow remote access to nodes that are part of different tree pieces.\cite{menon}   Communication across processors involves a   network.
	
\begin{figure}[h!]
\centering
\includegraphics[width=2in]{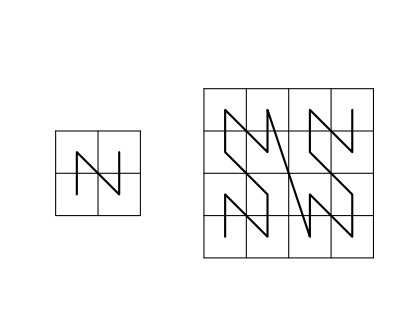} 
\caption{The left-hand side  is a square divided into four sub-squares with the ``curve'' going through each of the four sub-squares.  The curve looks like the letter ``N,'' which might not look like a  curve.  The figure is  helpful in understanding the meaning of a space-filling curve, because we can   see the interpretation of the  ``space'' that is required to understand the acronym ``space-filling curve.'' That is, there are only four ``points'' -- instead of an infinite number -- in this space.  The right-hand figure is  intended to help generalize the point of the left-hand subfigure. The figures are  from Ref.~\onlinecite{spacefilling}.}
\end{figure}
	
\section{Boundary Conditions and Ewald Sums}\label{sec:boundary conditions}

We know that the observable universe spans a distance of 13+ billion light years. Even very large simulations such as those performed as part of the IllustrisTNG project\cite{illustris} have computational volumes with widths only in hundreds of millions of light years.  Thus, even exceptionally large simulations model regions which   constitute only a small fraction of the universe. Because gravity  acts at large distances, this limitation is a  problem for realistic simulations. Before discussing an approach to this problem, we first discuss an important concept  of the currently accepted cosmological model: the cosmological principle. The cosmological principle states that the universe is both  isotropic and homogeneous.\cite{bob} Being isotropic means that the universe looks the same in all directions. Similarly,  homogeneous means that the universe looks the same at all locations. Readers who have looked at the night sky may object, because  the universe does not look the same in all directions and there are distinct features such as  stars and constellations. Similarly, others might argue that the universe must look   a little different for someone  in the Andromeda galaxy. The cosmological principle  needs to be understood in the context of distance scales: on some scales the universe is isotropic and homogeneous, but on others it definitely  is not.   The cosmological principle does not state that on a   scale such   as our observational perspective, the universe cannot contain distinct features, but  rather that at sufficiently large scales (relative to our perspective) the two assumptions hold.\cite{uoregon-cosmology}	

The cosmological principle provides the justification for modeling a simulation volume as a single cell in an infinite (or at least very large) three-dimensional grid of perfectly identical cubes,  which is  the approach taken by ChaNGa. Such codes use periodic boundary conditions, because the structure of the simulation volume is repeated over a periodic lattice.\cite{stadel} This approach addresses the issues   that arise from a finite simulation volume, but   it is  then necessary    to  treat the gravitational force calculation over an infinite lattice, which might seem no less daunting. The solution adopted by ChaNGa is to break up the calculation into long-range and short-range components.\cite{mitch,stadel} 
The short-range calculation is performed as an extension of the Barnes-Hut algorithm described in Sec.~\ref{sec:barnes-hut} by including a number of neighboring lattice cells (usually 26 of them\cite{stadel}) in the Barnes-Hut force calculation. The long-range gravitational contribution is then accounted for by Ewald sums,\cite{menon,stadel,ewaldsum} where the simulation cell is repeated exactly in every direction so that the long range force can be calculated  as a sum over all these repeated cells.
	
\section{Force Softening}\label{sec:force softening}
	
Another issue related to the gravitational force calculation is how to handle the gravitational force between particles at very small distances. From   Eq.~(\ref{eq:gravity vec}) we can see that  the interparticle force increases  very quickly at small particle separations. This infinite force  for zero separation  presents a problem because it can be difficult   to handle very large forces computationally,\cite{beckett}  leading to unphysical results.\cite{stadel} Hence, it is important for an N-body code to implement some kind of softening to impose a limit on the magnitude of the force. In ChaNGa this force softening  is closely related to the handling of  smoothed particle hydrodynamics (see   Sec.~\ref{sec:sph}) and involves   a spline softening kernel,\cite{menon,stadel} which effectively cuts off the gravitational force at zero interparticle separation, while maintaining the usual Newtonian gravity  at distances greater than  the softening length.\cite{wadsley2003} 

\section{Smooth Particle Hydrodynamics}\label{sec:sph}
	
We see that a robust and efficient computational treatment of gravity is no simple task. There remains several  nuances which have  been discussed only  superficially or  have been omitted entirely.\cite{stadel,menon,wadsley2003,wadsley2017} We could   write a  book just on the workings of a single sufficiently advanced N-body code.\cite{stadel} 
	Nevertheless, there is another crucial element of ChaNGa that we would be  remiss to not discuss at least briefly: gas dynamics. 
	
Everything we have discussed about ChaNGa so far has been related to predicting the motion of infinitesimal point masses under the influence of a mutual gravitational force. Clearly this calculation is crucial for computational astrophysics: gravity is a dominant force in the cosmos. However, simulating the motion of  particles under the influence of gravity  would  be sufficient only if we lived in a universe that consisted purely of discrete, massive bodies such as stars and dark matter. In particular, a purely gravitational simulation would   ignore the significant amounts of gas and dust present in our universe, which also play a significant role in galaxy formation and cosmological structure.\cite{beckett} Gravity  influences the motion of gas and dust, but additional considerations need to be made for the dynamics of the gas itself. To handle gas physics, N-body codes turn to fluid mechanics and treat the gas  as a continuous fluid medium, a reasonable approximation at a macroscopic scale.\cite{beckett,fluid}  The fundamental equation of fluid dynamics is the nonlinear Navier-Stokes equation.  This nonlinearity    leads to many more challenges.
	
There exist two approaches to fluid dynamics: the Eulerian description and the  Lagrangian description.\cite{fluid} The Eulerian description considers what is happening at fixed locations (or points) in space as a fluid medium flows through these points. In contrast,  the Lagrangian description considers the dynamic evolution of individual pieces of matter representing ``fluid elements'' as they travel through space.\cite{fluid}  Broadly speaking, there are two types of  N-body simulation codes that treat gas dynamics: grid-based (also known as  mesh-based) codes, which follow the Eulerian description by considering a simulation grid and tracking the flow of gas through that grid, and particle-based codes, such as ChaNGa, which follow the Lagrangian description by tracking the spatial motion of individualized gas ``parcels.''\cite{wadsley2003} 
		

One of the advantages to the Lagrangian approach is that it is a fairly natural extension of the Barnes-Hut based gravitational force calculation described in Sec.~\ref{sec:barnes-hut}.\cite{menon} The particles involved in the gravitational force calculation are also used to discretize a gas, effectively representing the ``parcels'' for the Lagrangian approach.\cite{beckett} This important and elegant step is accomplished through a process known as smoothed particle hydrodynamics, which uses the particles present in the simulation to derive continuous fluid quantities (such as the pressure and temperature) spanning the surrounding region of space.\cite{beckett,mitch} In this sense, the particles can be thought of as being ``smoothed out''  over space, smudged into a continuous particle-fluid that represents the gas. 
	
\begin{figure}[t]
\centering
\includegraphics[scale=0.4]{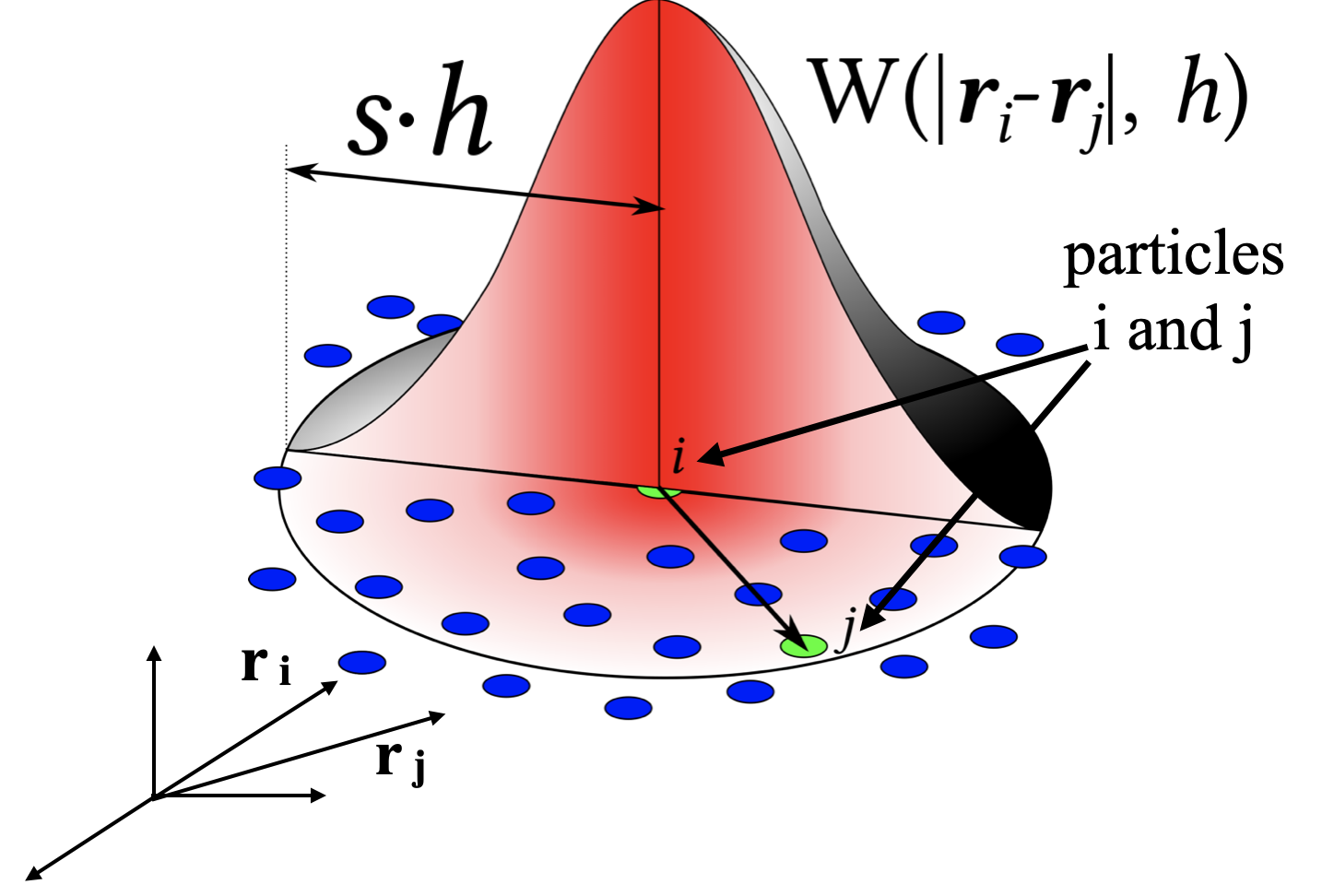}
\caption{(Color online) Depiction of a smoothing kernel $W$  applied to a discrete set of particles. The blue-filled circles represent the particles, $h$ the smoothing length, and the  curve represents the value of the smoothing kernel for all points in space. The frame of reference in the lower left is intended to help  understand the arguments of  $ W $.  The vector $ \vec{r}_{i} -  \vec{r}_{j }$  is shown in the plane of the blue-filled circles.\cite{wiki}}
\label{fig:sph}
\end{figure}
	
We  briefly provide a more technical description of the smoothing method used for smoothed particle hydrodynamics. A more detailed description is given in  Sec.~5 of Ref.~\onlinecite{wadsley2003}.  ChaNGa's implementation of smoothed particle hydrodynamics follows Ref.~\onlinecite{wadsley2003}. The method involves using a smoothing kernel to continuously map fluid quantities (provided by the representative particles) to the region of space around the particles.\cite{beckett} The smoothing calculation is given by\cite{mitch}
	\begin{equation}\label{eq:smooth}
	    \langle f(\vec{r})\rangle = \int f(\vec{r}_j)W(|\vec{r}_i-\vec{r}_j|,h)dV(\vec{r}_j),
\end{equation}
where $f(\vec{r}_j)$ is the fluid quantity (such as the density)  being mapped onto space for a representative particle, $\langle f(\vec{r})\rangle$ is its average value computed at position $\vec{r}$, $W$ is the smoothing kernel function, and $h$ is the smoothing length.\cite{beckett} Equation~(\ref{eq:smooth}) can  be approximated as  a sum over particles\cite{mitch}
\begin{equation}\label{eq:smooth sum}
	    \langle f(\vec{r}_i)\rangle=\sum_j V_jf_jW(|\vec{r}_i-\vec{r}_j|,h).
\end{equation}
A depiction of the smoothing process is shown in Fig.~\ref{fig:sph}. 	
\section{Results}\label{sec:results}
	
So far we have introduced N-body hydrodynamic simulation methods at the conceptual level because most researchers do not write code from scratch.  In the following, we give a brief overview of the user's perspective when running a simulation in ChaNGa on a laptop or desktop computer.  Running a ChaNGa simulation on a supercomputer requires considerably more infrastructure.\cite{XSEDE}
	
Given that you have access     to a computer running Linux,    doing a simulation in ChaNGa requires only  the code itself, a parameter file, and an initial conditions file. ChaNGa is a  complicated piece of software, and as such is supported by a number of dependencies.
	This complexity, as well as that of its dependencies, can  lead to some difficulties in the installation process.  For those having trouble installing the software, see Ref.~\onlinecite{changa-github} and   the supplemental material which provides some detailed notes on ChaNGa installation.\cite{lucas-first-summer}
	
ChaNGa simulates the gravitational motion and gas dynamics of matter in space, with the goal of modeling galactic and/or cosmological structure. The initial conditions file is a binary file (not human readable) that provides the initial positions and velocities of all particles in the simulation.\cite{mitch} These initial conditions  cannot be arbitrarily assigned and must be computed in a manner that approximates observations of the early universe via the  cosmic microwave background  measured by the Wilkinson microwave anisotropy probe.\cite{WMAP,planck} For a discussion of the generation of initial conditions for the AGORA project using MUSIC,\cite{music}  see Chapter 1 of Ref.~\onlinecite{mitch}.  A  more elementary and  well-documented program for producing  initial conditions that works well with ChaNGa  is pyICs.\cite{herpich}  The initial conditions   provide the starting point    by giving the initial position and velocity values for all the particles; these values are assigned to be approximately consistent with observation.
	
\begin{figure}[t]
   \centering
   \includegraphics[width=5.5in]{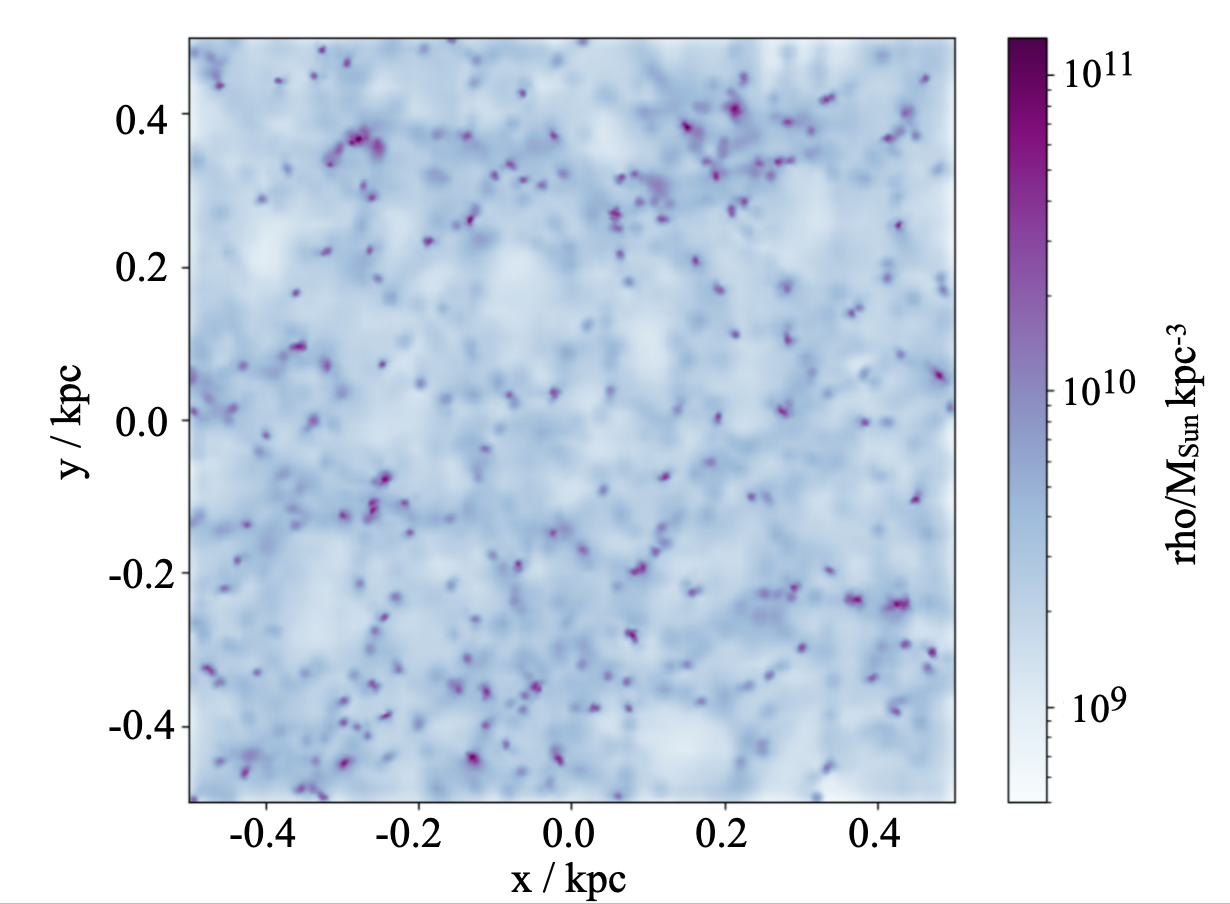} 
   \caption{A visualization of the  cosmos at redshift zero based on a cube300 test simulation in the ChaNGa directory. (Reference~\onlinecite{supplmentalmaterial}  provides details for running cube300.)  Note the web-like pattern of the density: in many regions the density is zero.  The  fibril-like paths at densities indicated by light blue are often called filaments.}
   \label{fig:testcosmo-4}
\end{figure} 

We used the  software	visualization tool \texttt{pynbody}\cite{pynbody} to generate the image in Fig.~\ref{fig:testcosmo-4} based on the cube300 simulation.\cite{supplmentalmaterial} Pynbody is suggested  because it is  particularly ChaNGa friendly, and the documentation is excellent. The fibrils of density corresponding to light blue are of great importance for transporting gas to and from galaxies and their physics is  of great interest.\cite{nir}  The purple dots in the filament represent galaxies.
	
Our goal has been to extend the presentations of the N-body hydrodynamic code ChaNGa given in Refs.~\onlinecite{changa-github, menon} to make using ChaNGa more accessible.  Some advanced topics at the heart of ChaNGa have regretfully been left out.  It is hoped that others with experience using ChaNGa might take up the challenge of providing a part two on possible topics such as the CHARM++ language and how it works with various kinds of hardware.

\section{Suggested problems}\label{sec:exercises}

A few  suggested problems are provided to guide the reader in
both the analysis of the simulation results and  basic concepts 
in  cosmological physics.
\bigskip

\bigskip \noindent {\it Problem 1}. Units in ChaNGa are interesting and require attention.  Provide a detailed derivation of the results for mass units in  ChaNGa wiki example number two under ``Understanding Units,'' which is under ``Code Units.''\cite{testcosmo}   The derivation will get the reader ready to understand the units and value of the Hubble constant in Problem~2. Reference~\onlinecite{tipsy} is extensive and requires some study before  finding ``Code Units.'' 

\noindent {\it Problem 2}. Imagine that you are responsible for creation and that one of your tasks
is to create a multiverse.  Simulate a universe that has a Hubble constant $H_0$ that is, say, roughly half the value of the presently accepted value.  Use ChaNGa and cube300 to generate a 
$ H_{0}/2 $ universe and demonstrate that the  simulation behaves as expected. Note: this problem is a bit -- possibly very -- tricky, but once the data is found/obtained and compared, the result is very clear. 

This problem depends  strongly on understanding the material in the supplemental material \cite{supplmentalmaterial} For example,  The supplement  provides the Hubble Constant parameter and explains the issue of initial conditions  for this simulation.

\bigskip \noindent {\it Problem 3}. Use Pynbody to demonstrate that, in fact, cube300 is a dark matter only simulation.  The installation documentation for Pynbody from GitHub is quite good.\cite{nir} Some knowledge of Python 3 is necessary to use Jupyter to access the low-level abilities of Pynbody for the solution to this problem.

\begin{acknowledgments}
We gratefully acknowledge Reed College for summer research internship support. J.P. would also like to acknowledge ACCESS allocation AST200020 for supporting this project.
\end{acknowledgments}

\newpage

\begin{center}
\textbf{Supplementary Material \# 1}
\end{center}

\begin{center}
\textbf{ The User's Perspective} 
\end{center}

\begin{center}
J. W. Powell, L. Caudill, and O. Young
\end{center}

\begin{center}
Reed College, Department of Physics, Portland, Oregon 97201
\end{center}


\begin{center}

ABSTRACT

In Ref.~\onlinecite{AJParticle} we  introduced  N-body hydrodynamic simulation methods at the conceptual level because most researchers need to understand the background of the code and only a few researchers wrote  ChaNGa from scratch.  In this supplemental document number 1, we give a brief overview of the user's perspective when running a simulation in ChaNGa on a Linux laptop or desktop computer that augments the github wiki at Refs.~\onlinecite{changa-github}.  Running a ChaNGa simulation on a supercomputer requires considerably more infrastructure.\cite{XSEDE} Please note that the figure numbers for this supplementary document follow the figure numbers for the main document instead of having new numbers.

\end{center}

\maketitle


\noindent \textbf{I. THE USER'S PERSPECTIVE} 
\bigskip
	
Given that you have access     to a computer running Linux,    doing a simulation in ChaNGa requires only  the code itself, a parameter file, and an initial conditions file. The last two files are explained in the following. ChaNGa is an incredibly complicated piece of software, and  is supported by a number of dependencies. This complexity, as well as that of its dependencies, can  lead to  difficulties in the installation process.  For those having trouble installing the software, see Refs.~\onlinecite{changa-github} and  \onlinecite{lucas-first-summer}.
	
ChaNGa simulates the gravitational motion and gas dynamics of matter in space, with the goal of modeling galactic and/or cosmological structure. The initial conditions file is a binary file (not human readable) which provides the initial positions and velocities of all the particles in the simulation.\cite{mitch} These initial conditions  cannot be arbitrarily assigned and must be computed in a manner that approximates observations of the early universe via the  cosmic microwave background  measured by the Wilkinson microwave anisotropy probe and more recently the Planck probe.\cite{WMAP,planck} For a discussion of the generation of initial conditions for the AGORA project using MUSIC,\cite{music}  see Ref.~\onlinecite{mitch},  Chap.~1.  A  more elementary and  well-documented program for producing  initial conditions that works well with ChaNGa  is pyICs.\cite{herpich}  The initial conditions   provide the starting point for the computation by giving the  positions and velocities for all the particles; these values are assigned to be approximately consistent with observation.
	
\begin{figure}[t]
\centering
\includegraphics[scale=0.6]{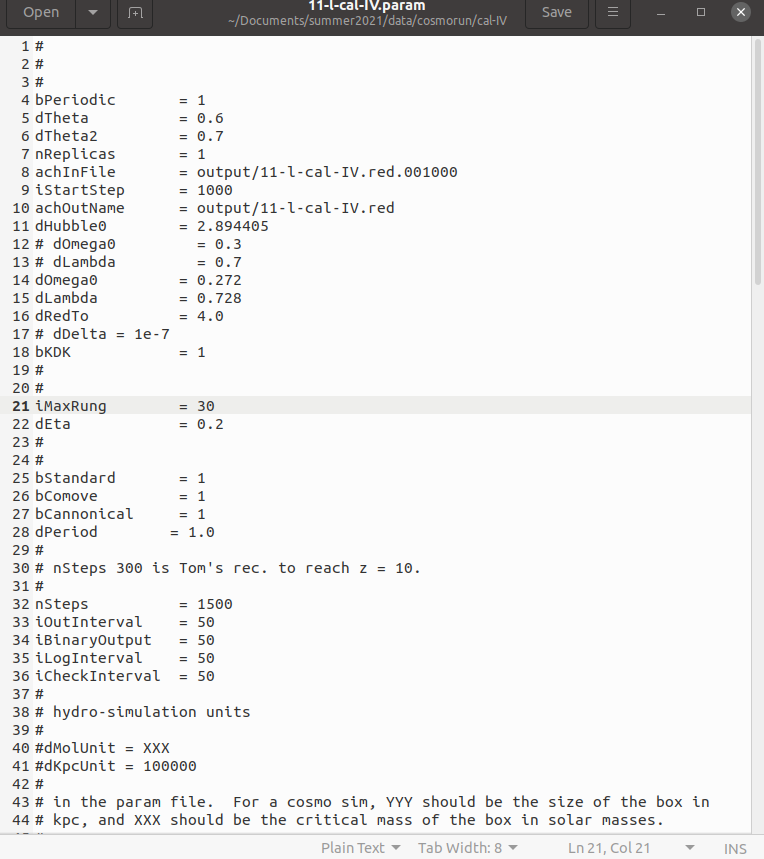}
\caption{The parameter file for ChaNGa from Ref.~\onlinecite{parameter}.  There are at least two kinds of restarts in ChaNGa: from a checkpoint file and from a snapshot.  The right-hand side of line~8 is the path to a snapshot from which the simulation restarted. The initial conditions file is not present  on line~8,  and this example   parameter file is a  restart from an earlier simulation.}
\label{fig:param}
\end{figure}
	
The parameter file shown in Fig.~\ref{fig:param} is a human readable file with a \texttt{.param} extension that specifies a number of parameters and important options.\cite{changa-github-wiki} The parameters (except for a few notable exceptions) are named such that their data type is specified by their first letter: \texttt{b} for Boolean, \texttt{i} for integer, and \texttt{d} for double precision. Some of the  many options worth noting are \texttt{dLambda}, which represents the cosmological constant $\Lambda$ (see Ref.~\onlinecite{bob}, Chap.~29), \texttt{dHubble}, which represents the Hubble Constant $H_0$,\cite{bob} \texttt{nSteps},  which gives the number of time steps to be performed, and \texttt{iOutInterval}, which gives the number of time steps between  outputs. One place to begin  understanding  the ChaNGa options for the parameter file is Ref.~\onlinecite{changa-github-wiki}. Parameters are chosen with physical realism and/or computational efficiency in mind. 
	
Once the user has a working version of ChaNGa, an initial conditions file, and a parameter file, performing the simulation  involves a  terminal command with many components specifying the number of processor cores to use, as well as the location and name of the parameter file (the name of the initial conditions file is specified as the \texttt{AchInFile} parameter).  An understanding of the Linux command line\cite{no-starch-press-linux-com-line} is an important component of using ChaNGa.  To run the program on a supercomputer, a \texttt{dot sh} file is required     to  schedule   the requests.\cite{slurm}
\bigskip


\noindent \textbf{II. A FIRST SIMULATION} 
\bigskip
	
Reference~\onlinecite{testcosmo}  has a section on suggested tests  or first simulations.   The testcosmo simulation produces excellent data which corresponds  well with reproducible and well-established experiments.  Here we augment Ref.~\onlinecite{testcosmo}. 
	
\begin{figure}[t]
   \centering
   \includegraphics[width=6in]{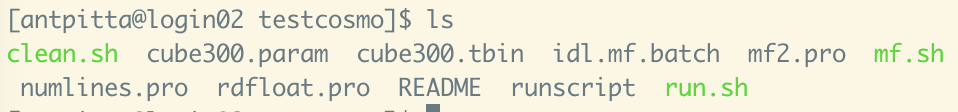} 
   \caption{A screen dump of the testcosmo directory, a subdirectory of the changa directory.  The commend \texttt{ls}  lists the files in a directory.  The green text indicates a script file. The cube300.tbin file is the initial conditions file, and tbin stands for tipsy binary.}
   \label{fig:testcosmo-1}
\end{figure}

Once ChaNGa has been installed, the first task is to find the testcosmo directory.  The testcosmo directory is found in the same directory that has the changa executable and many of the other header files and C++ like files.   The testcosmo directory  includes the  list of files shown in Fig.~\ref{fig:testcosmo-1}.  Only a few of these files are needed to run  an  interesting simulation. The second file, moving from left to right, cube300.param, is particularly important because it is an introductory example of a parameter file.  The letters ``cube'' refer to the fact that this is a cosmological simulation in the sense described in Sec.~VI of Ref.~\onlinecite{AJParticle}. 

\begin{figure}[t]
   \centering
   \includegraphics[width=6in]{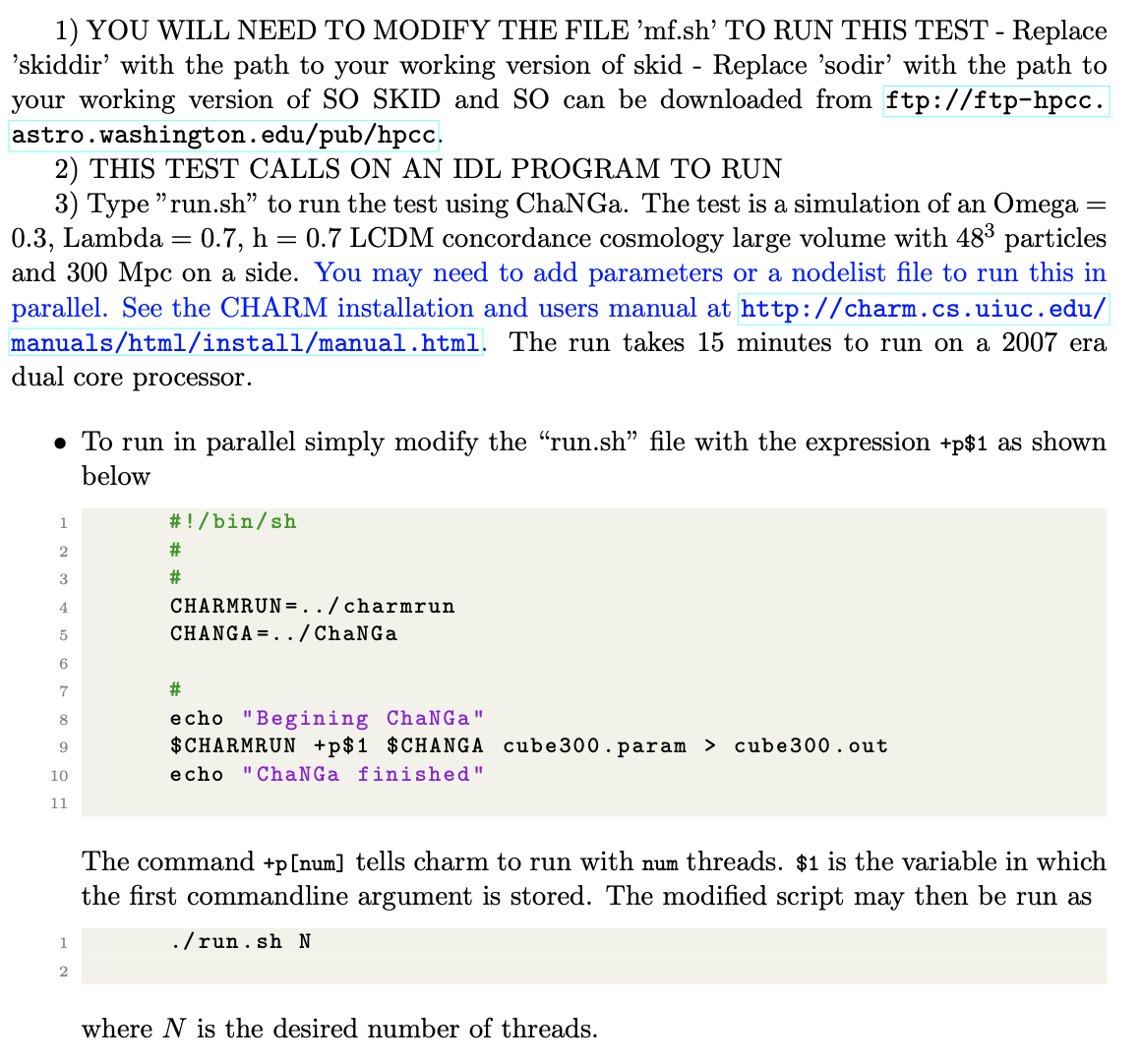}
   \vspace{-0.2in}
   \caption{ The first three sections of the testcosmo README file.  The first two sections may be skipped for the purposes of this paper. The third section has been modified by us to discuss how we implemented the parallelization of testcosmo. Those unfamiliar with the concept of threads are referred to  Ref.~\onlinecite{threads}. }
\label{fig:testcosmo-2-b.png}
\end{figure} 

\begin{figure}[t]
   \centering
   \includegraphics[width=6in]{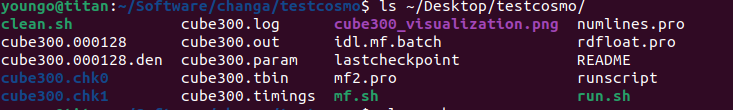} 
\caption{A screendump of the testcosmo directory after a successful run of the cube300 cosmological simulation. There are now several more files, e.g., \texttt{cube300.log}, \texttt{cube300.timings}, \texttt{cube300.chk0}, \texttt{cube300.chk1}, and \texttt{lastcheckpoint}, which are  important in various contexts but beyond the scope of this introduction.  The  important file is \texttt{cube300.000128} because it is the data output of the simulation.  We used  this file to make the image of the cosmos in Fig.~6 in Ref.~\onlinecite{AJParticle}. }
   \label{fig:testcosmo-3.png}
\end{figure}

The file \texttt{cube300.tbin} represents a tipsy binary file\cite{tipsy} and contains the initial conditions. (Note: these initial conditions cannot be easily altered.  For the purposes of Problem~2 in Ref.~\onlinecite{AJParticle}, the cube300 initial conditions should be assumed and are a ``black box.'' The interested reader can consult  Binney and Tremaine. \cite{binnney-tremain} Recall from the Main document that the program MUSIC can generate initial conditions.). The first three sections of the README file  in Fig.~\ref{fig:testcosmo-2-b.png} are composed of both text and command line screen dumps.  The last two sections of the README file (not shown) provide fascinating information about extending the interpretation and understanding of the physics of the data produced by the testcosmo simulation. It is possible to bypass points 1 and 2 of the README and produce a simulation. The third section is crucial, especially the first sentence which introduces the \texttt{run.sh} file.  This file can be read and modified by a command line text editor. The second sentence provides the physical data for the simulation and is designed for someone with a strong understanding of cosmology. The third section explains the ``300'' in cube300. We have added a section to the README to explain our method of running the cube300 simulation in parallel.

Figure~\ref{fig:testcosmo-3.png} shows a screen dump of the testcosmo directory after completion of a simulation.   There are many more files than before the simulation, i.e., in Fig. \ref{fig:testcosmo-1}, but the most important one is  \texttt{cube300.000128}.  The number 000128 indicates  the number of the time steps for which data was produced.  The data in this  file   can be visualized  using several kinds of visualization software.

\newpage

\begin{center}
\textbf{Supplementary Material \# 2}

\end{center}

\begin{center}

\textbf{Installing ChaNGa}

\end{center}


\begin{center}
 
J. W. Powell and L. Caudill

Reed College, Department of Physics, Portland, Oregon 97201

\end{center}



\begin{center} 

ABSTRACT


    This document outlines the process used to install ChaNGa on an Ubuntu computer as performed in May of 2020, and provides some tools that might be of use in, say, 2023. This process is derived from the information outlined in Section 1 of \textit{Summer 2019 N-body Astrophysics Simulation Research Overview} by W. Lum, B. Cummings, and J. Powell, which is available upon request from the corresponding author.
    It is likely that future versions of ChaNGa and/or Charm++ will require broad interpretation of these instructions. 
    Given the often challenging nature of installation, however, the reader is encouraged to consider the suggestions herein that worked.
    As in the first Supplementary document, the figure numbers for this supplementary document follow the figure numbers for the main document instead of having new numbers.
\end{center}
\bigskip



\noindent \textbf{I. INTRODUCTION} 
\bigskip

This document is intended for an undergraduate physics student who has recently completed their sophomore year. In writing the document we have tried to make as few assumptions as possible regarding the level of computer science proficiency of the reader. Nevertheless, completing the installation process requires knowledge of basic Linux terminal navigation and knowledge of the website GitHub, neither of which are covered in most first and second year physics classes at Reed College. Readers who are new to Linux
are enthusiastically encouraged to study Ref. 41 of the corresponding AJP article before attempting
an installation of ChaNGa. 
\bigskip


\noindent \textbf{II. CHARM++} 
\bigskip

\subsection{Installation and Setup}
\begin{figure}[h]
    \centering
    \includegraphics[scale=0.3]{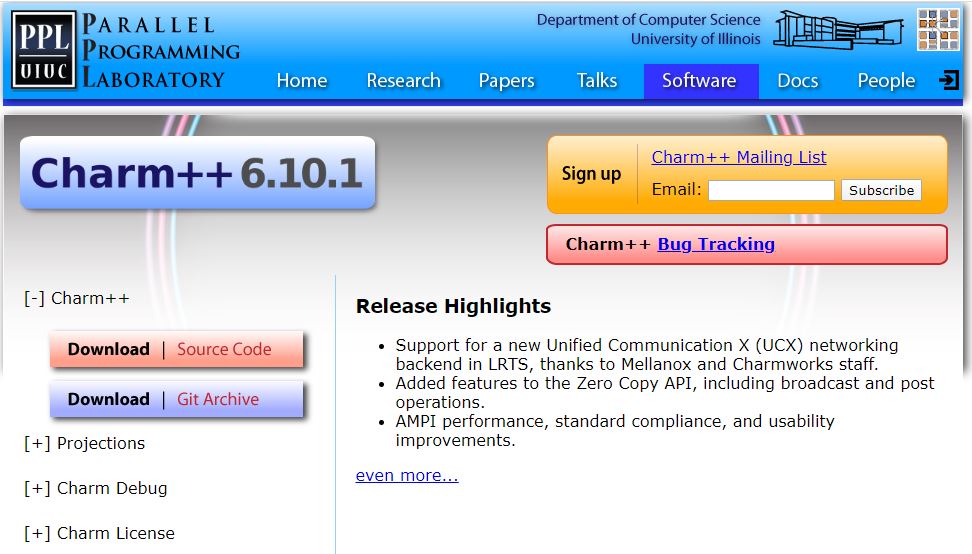}
    \caption{The Charm++ webpage, as of May 2020.}
    \label{fig:charm}
\end{figure}

The first step to installing ChaNGa, as described on the github wiki \cite{changa} is to install Charm++ from the University of Illinois at Ref. \cite{charm}. Note that, as per the recommendation of \cite{summer2019}, version 6.8.2 of Charm++ should be installed rather than the latest release to avoid issues when compiling ChaNGa. As of May 2020, this can be done by clicking the ``Download Source Code" button on the left side of the page (see Fig.~\ref{fig:charm} 
) and selecting ``Charm 6.8.2 Source Code". Similar kinds of actions maybe required for successful installation of ChaNGa at anytime, e.g. 2023. This download will tarball a file containing the desired version of Charm++ to your computer.

Once installed, the tarball file will need to be extracted, and the resulting directory will need to be moved to the desired location within the computer's file system. The easiest way to extract the tarball is to use your computer's user interface to navigate to the Downloads folder and right-click on the file, selecting the option to extract. Alternatively, the tarball can be extracted from the Linux terminal if the user navigates to the 
\texttt{Downloads} directory and inputs the following command:
\[\texttt{\$ tar -xvf [filename]}\]


Extracting the tarball should add a directory named 
\texttt{charm-[version]} (in this case, 
\texttt{charm-6.8.2}) to the \texttt{Downloads} directory. (The current version of Charm++ is 7.0.0 -- 2023-feb-03.) At this point, the user needs to rename this directory to ``charm" and move it elsewhere within the directory tree. It is recommend doing both simultaneously using the 
\texttt{mv} command, for instance, the command
\[\texttt{\$ mv \texttt{charm-[version]} $\sim$/charm}\] 
will rename the directory and move it to \texttt{/home/[user]/}. As described in \cite{changa}, the user should then navigate to the charm directory and enter the command below that needed to be spread onto two lines:
\[\texttt{\$ ./build ChaNGa netlrts-linux-x86}\]
\[\texttt{\_64 --with-production}\]
to build charm with the appropriate ChaNGa libraries. 



\bigskip

\noindent \textbf{III. CHANGA} 
\bigskip

If you've made it this far, you likely have a working version of charm++. Congratulations! The next step is obtaining and setting up ChaNGa. 

First, you should navigate to the directory in your file system that contains the \texttt{charm} directory that was installed in Section~\ref{sec::charm}. This suggestion means that when you enter the terminal command \texttt{ls}, \texttt{charm} should be listed as a subdirectory. Starting from directory that contains \texttt{changa} ensures that you need not set the environment variable \texttt{CHARM\_DIR}, as described in \cite{changa}. It should be noted that if you are comfortable with setting the environmental variable, then you can put ChaNGa anywhere in your file system, regardless of the location of \texttt{charm}.

Once you've navigated to the directory where you intend to install ChaNGa, enter the following commands (as per the instructions on \cite{changa})\    \bigskip 

    \texttt{\$ git clone https://github.com/
    N-BodyShop/changa.git}
    \bigskip 
    
     \texttt{\$ git clone https://github.com/
    N-BodyShop/utility.git}
    \bigskip 
    
\begin{figure}[h]
    \centering
    \includegraphics[scale=0.5]{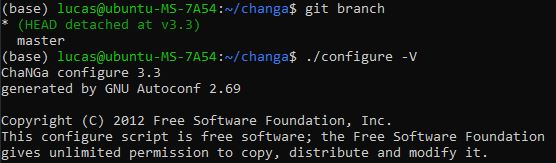}
    \caption{The terminal outputs of \texttt{git branch} and \texttt{./configure -V} after successfully switching to ChaNGa version 3.3}
    \label{fig:changa_version}
\end{figure}

These commands will create two new subdirectories, \texttt{changa} and \texttt{utility}, 
in your current directory. 
You will now need to switch to ChaNGa version 3.3 \cite{summer2019}. This can be done by navigating to the \texttt{changa} directory and entering the command 
\[\texttt{\$ git checkout v3.3}\]
You may consider running the \texttt{configure} file with the command
\[\texttt{\$ ./configure}\]
To ensure that the change in version is applied. You can verify that you've successfully switched versions by entering
\[\texttt{\$ ./configure -V}\]
Additionally, you can use the command 
\[\texttt{\$ git branch}\]
to check that you've successfully switched to the v3.3 GitHub branch of ChaNGa.
These two commands should produce terminal outputs similar to those displayed in Fig.~\ref{fig:changa_version}.

At this point in 2020, attempting to compile ChaNGa with the \texttt{make} command was likely produce a compiler error reporting that the variable \texttt{HUGE} in the file \texttt{InOutput.C} is undefined. In ChaNGa v3.3, this variable should be present in lines 55 and 568 of \texttt{InOutput.C} (though it may be worth double checking the compiler error to make sure this is the case).

The solution to this compiler error, as documented in \cite{summer2019}, is to edit the \texttt{InOutput.C} file using your preferred text editor, going to each line containing the variable \texttt{HUGE} and replacing it with \texttt{FLT\_MAX}. 
Doing so should resolve the compiler error. Enter the command
\[\texttt{\$ make}\]
If successful, this will add the executables \texttt{charmrun} and \texttt{ChaNGa} to the \texttt{changa} directory, and you should have a working version of ChaNGa. You can test your version of ChaNGa by navigating to the \texttt{testcosmo} directory and running the tests documented on the "Running ChaNGa" section of the ChaNGa wiki at \cite{changa}.


\begin{thebibliography}{99}
\bibitem{wiki}The image in Ref.~\onlinecite{fig:sph} was uploaded to Wikipedia by   Jlcercos and is licensed under Creative Commons.
\end{thebibliography}
\end{document}